\documentclass[sigconf, nonacm]{acmart}

\newcommand\vldbavailabilityurl{}

\usepackage{booktabs}

\begin{document}
\title{Exploring the Semantic Gap in Agentic Data Systems: A Formative Study of Operationalization Failures in Analytical Workflows}

\author{Jalal Mahmud}
\affiliation{%
  \institution{Megagon Labs}
  \city{California}
  \state{USA}
}
\email{jalal@megagon.ai}

\author{Eser Kandogan}
\affiliation{%
  \institution{Megagon Labs}
  \city{California}
  \country{USA}
}

\email{eser@megagon.ai}

\begin{abstract}

Large language models (LLMs) are increasingly used to generate
queries, invoke tools, and construct analytical workflows.
Although recent advances have substantially improved workflow
generation and execution, the semantic information required to
operationalize analytical concepts often lies beyond what is
explicitly represented in database schemas and data values. We present a cross-domain formative study of operationalization
failures in agent-generated analytical workflows. Across 236
analytical intents spanning finance, human resources, and public
safety domains, we identify 153 recurring failures despite
successful workflow generation and execution. Our analysis
reveals five recurring classes of failures: comparative grounding,
process reasoning, quantitative reasoning, role confusion, and
policy grounding. These findings suggest a semantic gap between user-level
analytical concepts and the information available to
workflow-generation systems. More broadly, they raise questions
about the admissibility of analytical operations and suggest that
future agentic data systems may require richer semantic
representations to bridge the gap between analytical intent and
executable computation.

\end{abstract}

\maketitle


\ifdefempty{\vldbavailabilityurl}{}{
\vspace{.3cm}
\begingroup\small\noindent\raggedright\textbf{VLDB Workshop Artifact Availability:}\\
The source code, data, and/or other artifacts have been made available at \url{\vldbavailabilityurl}.
\endgroup
}


\section{Introduction}

Large language models (LLMs) are rapidly becoming a core primitive of modern data systems. Modern AI-native systems increasingly rely on LLMs to generate queries, retrieve evidence, invoke tools, and construct analytical workflows. Examples include NL2SQL systems, retrieval-augmented systems, enterprise analytical assistants, and agentic planners that combine reasoning with database and retrieval operations \cite{pourreza2023dinsql,pourreza2025chasesql,liu2025xiyansql,khattab2024dspy,Kandogan2025Blue}.
Recent advances have substantially improved intent interpretation,
workflow generation, and answer correctness. Nevertheless,
analytical failures continue to occur even in workflows that
execute successfully.

Consider an analytical assistant tasked with answering the
request: \emph{`Find unusually risky loans with sustained
delinquency.''} A generated workflow may operationalize
\emph{`unusually risky''} using a fixed threshold and
\emph{`sustained delinquency''} using a row-level predicate.
However, unusual risk is inherently comparative and typically
requires reasoning relative to a reference population, while
sustained delinquency describes a temporal process that cannot
be inferred from a single observation. Similarly, when asked to \emph{`identify intersections with the
highest fatality rate,''} the generated workflow may rank
intersections using total fatalities rather than normalized
fatality rates. Such a workflow conflates counts with rates,
favoring high-volume intersections even when their fatality rate
is lower. In both cases, the generated workflow is syntactically valid and
executable, yet the selected operations do not faithfully capture
the analytical concept expressed in the user's intent.

\begin{figure*}[t]
\centering
\includegraphics[width=0.75\textwidth]{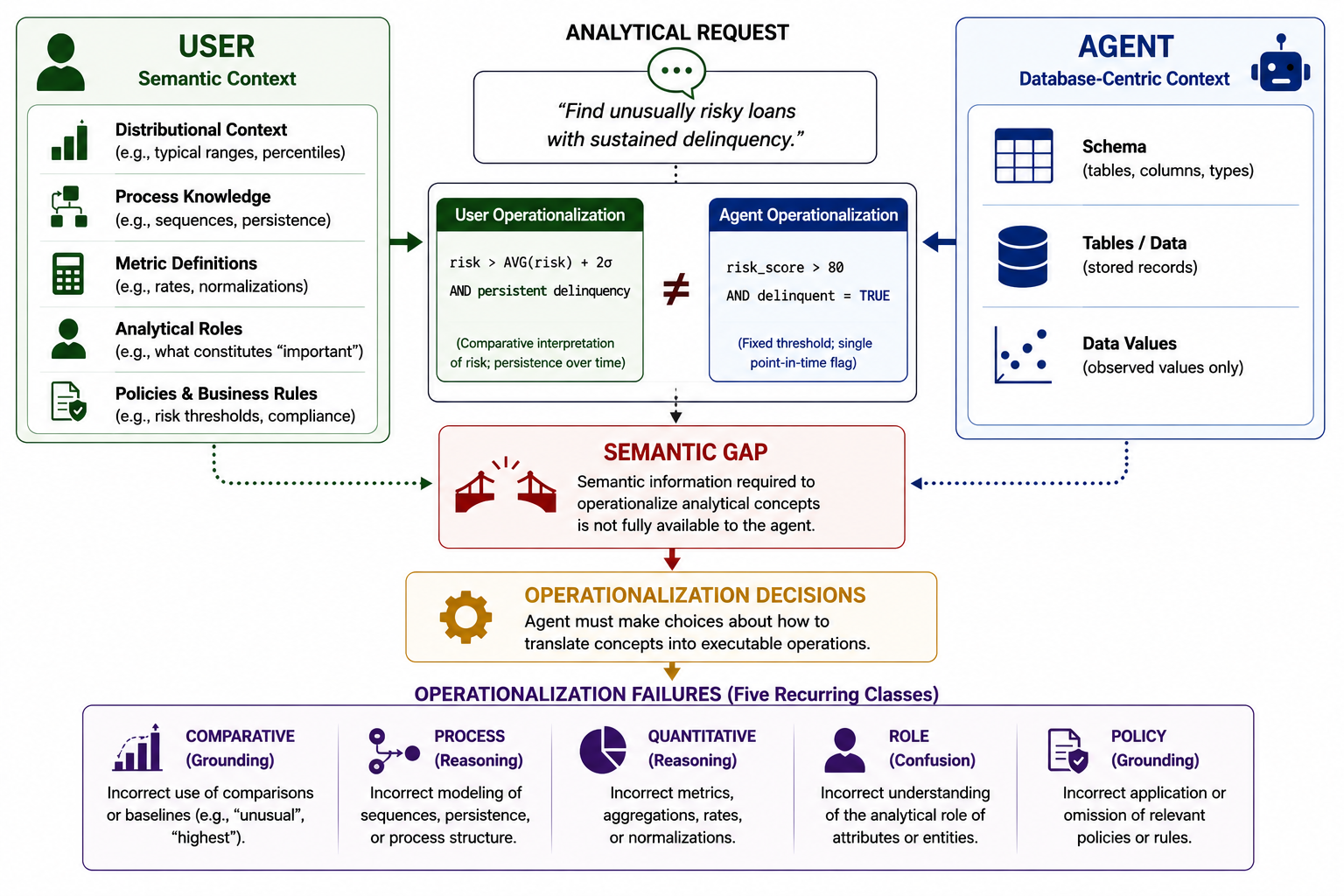}
\caption{Illustration of the semantic gap in agentic data systems. Information required to operationalize analytical concepts may not be fully available to workflow-generation systems, leading to operationalization failures despite successful execution.}

\label{fig:semantic_gap}
\end{figure*}

These examples suggest a broader semantic gap in agentic data
interaction. Existing data systems have traditionally focused on
bridging gaps in intent interpretation, schema understanding,
and query generation. Our findings suggest an additional
semantic frontier concerned with the operationalization of
analytical concepts. Concepts such as \emph{unusual},
\emph{persistent}, \emph{high-risk}, and \emph{rate} must
ultimately be translated into executable computations, yet the
information required to perform this translation is often not
explicitly represented in database schemas or data values.

As shown in Figure~\ref{fig:semantic_gap}, human analysts
routinely draw on statistical context, process knowledge,
metric definitions, analytical roles, and organizational policies
when interpreting such concepts, whereas agents often have
access only to database schemas, data values, and limited
metadata.

To better understand this phenomenon, we conducted a
cross-domain formative study spanning finance, human resources,
and public safety analytics. Across 236 analytical intents, we
identified 153 recurring failures despite successful workflow
execution. Analysis revealed five recurring classes of failures:
comparative grounding, role confusion, process reasoning,
quantitative reasoning, and policy grounding. Collectively, these findings suggest a broader concern regarding
whether the analytical operations selected by a workflow are
appropriate for expressing the intended analytical concept. We
refer to this concern as analytical admissibility and discuss its
implications for future AI-native analytical systems.

\section{Related Work}

\textbf{NL2SQL and Agentic Analytical Systems}
Recent advances in large language models have substantially
improved natural-language interfaces to structured data. Modern
NL2SQL systems leverage schema linking, retrieval augmentation,
in-context learning, self-correction, decomposition, and
candidate selection to generate increasingly complex analytical
queries~\cite{yu2018spider,li2024bird,pourreza2023dinsql,
pourreza2025chasesql,lei2025spider2}. More broadly, agentic
systems extend query generation to multi-step planning, tool
use, and workflow construction over heterogeneous data
sources~\cite{yao2023react,liu2025xiyansql,khattab2024dspy,
Kandogan2025Blue,zeighami2025proactive}. These systems have
significantly improved the ability to interpret user intent and
generate executable analytical workflows. 
Our work investigates
a complementary challenge: whether a generated query or
workflow faithfully captures the analytical concept expressed in
the user's intent once it has been generated and executed
successfully.

\textbf{Failure Analysis of LLM-Powered Systems}
Prior work has examined failure modes in LLM-powered and agentic
systems, including reasoning errors, planning failures, and tool-use
failures~\cite{yao2023react,shi2024agentbench,
yao2025taubench}.
Within NL2SQL, benchmarks such as Spider, BIRD, and Spider 2.0
expand evaluation from semantic parsing to database-grounded
reasoning and enterprise-scale workflow generation~\cite{yu2018spider,
li2024bird,lei2025spider2}. Recent work has also highlighted gaps
between benchmark performance and real-world analytical
usability~\cite{floratou2024nl2sql}. 
Our work is complementary to these efforts. 
Prior work primarily studies intent identification, schema linking,
reasoning, and query generation. In contrast, we investigate
failures that arise even after the intended query or workflow has
been generated and executed successfully. The challenge is not
whether a workflow can be produced, but whether the selected
analytical operations appropriately instantiate the intended
analytical concept.

\textbf{Semantic Representations for Data Systems}
Prior work in data management has explored schemas, metadata
catalogs, semantic layers, ontologies, business glossaries, and
data constraints as mechanisms for improving data interpretation
and governance~\cite{codd1993olap,kimball2013datawarehouse,
abedjan2015profiling}. Modern metadata platforms such as
DataHub~\cite{datahub}, OpenMetadata~\cite{openmetadata} and
Amundsen~\cite{amundsen}
enrich data assets with lineage,
governance metadata, and business semantics, while semantic-layer
technologies make business concepts and metrics explicit for
analytical applications.
More broadly, measurement theory studies which operations are meaningful for different classes of quantities~\cite{stevens1946measurement}. Recent work on schema-aware language models and table representation learning~\cite{yin2020tabert,herzig2020tapas, deng2020turl,yu2021grappa} learns semantic representations of tables, columns, and relational context to improve tasks such as semantic parsing, retrieval, and table question answering. These approaches improve contextual understanding of structured data through learned semantic representations.

Our work is complementary to these efforts. Prior work primarily focuses on representing, organizing, governing, and exposing semantic information that helps users and systems interpret data assets. In contrast, we investigate how semantic information is used during workflow generation to determine which analytical operations are appropriate for expressing a given analytical concept. The operationalization failures identified in our study suggest a complementary need for semantics that guide the translation of analytical concepts into executable computations and constrain which analytical operations are analytically admissible in a given context.

\begin{table*}[t]
\centering
\small
\caption{Recurring operationalization failure classes identified in the formative study. Percentages are computed over the 153 identified operationalization failures.}
\label{tab:failure_modes}

\begin{tabular}{p{0.21\textwidth} p{0.36\textwidth} p{0.27\textwidth} p{0.06\textwidth}}

\toprule
\textbf{Failure Class} &
\textbf{Description} &
\textbf{Illustrative Example} &
\textbf{\%} \\
\midrule

Comparative Grounding &
Comparative concepts are operationalized using arbitrary thresholds rather than distributional or domain-grounded interpretations &
``severe collisions'' $\rightarrow$ \texttt{injured >= 10}
&
28\% \\

Role Confusion &
Attributes are assigned inappropriate analytical roles, treating counts, frequencies, or identifiers as evidence for higher-level concepts &
``most demanding job titles'' $\rightarrow$
\texttt{ORDER BY COUNT(*)}
(frequency $\neq$ intensity)
&
14\% \\

Process-Level Reasoning &
Process-oriented states emerging from event sequences are reduced to row-level predicates, ignoring persistence, recurrence, or temporal evolution &
``sustained negative balance'' $\rightarrow$
\texttt{balance < 0}
&
26\% \\

Quantitative Reasoning &
Derived quantities are constructed without respecting aggregation scope, functional dependencies, normalization requirements, or measurement constraints &
``total amount paid back'' $\rightarrow$
\texttt{amount + payments * duration}
(principal counted twice)
&
13\% \\

Policy Grounding &
Institutionally defined concepts are operationalized using local heuristics or observed data statistics rather than externally defined criteria &
``reportable incident'' $\rightarrow$
severity threshold inferred from data
&
19\% \\

\bottomrule
\end{tabular}
\end{table*}

\section{Formative Study of Operationalization Failures}

To better understand how analytical failures arise in agent-generated workflows, we conducted a multi-domain formative study spanning finance, human resources, and public safety analytics.

\textbf{Study Setup} We selected three representative relational datasets covering finance, human resources, and public safety domains. The finance dataset was drawn from the BIRD benchmark~\cite{li2024bird}; the public safety dataset was derived from the NYC Motor Vehicle Collisions dataset publicly available via
Kaggle~\cite{kaggle_nyc_collision}; and the human resources dataset was constructed from anonymized enterprise data. 
All experiments were conducted using a GPT-4o-based NL2SQL
agent implemented within the Blue platform~\cite{Kandogan2025Blue}.
The agent generated analytical queries using schema metadata,
attribute descriptions, profiling statistics, and representative
sample values, without handcrafted reasoning rules or
domain-specific analytical guidance.
We constructed 250 natural language intents reflecting realistic
enterprise analytical tasks
in finance, human
resources, and public safety domains. After feasibility validation, 236
executable intents were retained.

\textbf{Failure Identification}
For each intent, the NL2SQL agent generated and executed a SQL
query. We manually inspected the generated query and its result.
A query was classified as an operationalization failure if it was
(1) syntactically valid, (2) executed successfully, and (3)
reflected a different analytical interpretation than the one
expressed in the natural language intent.
Failure categories were derived through iterative
qualitative analysis and each failure was assigned to a single
dominant class.

\textbf{Key Findings} Our analysis yielded three principal findings.
\paragraph{Finding 1: Operationalization failures are common.}
Across 236 analytical intents, we identified 153
operationalization failures despite successful query execution.
The remaining 83 intents were judged to be appropriately
operationalized. 

\paragraph{Finding 2: Operationalization failures generalize beyond the study workload.}
To assess external validity, we analyzed two independently
constructed workloads from the BIRD benchmark~\cite{li2024bird}.
First, we  analyzed 31 analytical intents from the BIRD
MiniDev financial benchmark~\cite{li2024bird}. 
Despite being
independently constructed, 19 of 31 intents (61\%)
exhibited
operationalization failures.
Second, we analyzed 145 intents labeled \emph{challenging} by the
BIRD benchmark authors~\cite{li2024bird}, spanning 11 databases
across diverse domains. Of the 130 intents that produced executable queries (15 were
excluded due to API timeouts or SQL generation failures), 88
(68\%) exhibited operationalization failures.
This
suggests that operationalization failures
are not specific to
our study workload and may arise more broadly in analytical
workflow generation tasks.

\paragraph{Finding 3: Operationalization failures are highly structured.}
Rather than arising from isolated model mistakes, the observed
failures clustered into a small number of recurring categories
that appeared consistently across domains and benchmarks.

\begin{table*}[t]
\centering
\small
\caption{
Illustrative examples of operationalization failures and
knowledge-aware alternatives. The observed operationalization may be executable, but lacks the semantic context needed to faithfully express the intended analytical concept.
}
\label{tab:operationalization_examples}
\begin{tabular}{p{0.18\textwidth} p{0.22\textwidth} p{0.25\textwidth} p{0.25\textwidth}}
\toprule
\textbf{Failure Class} &
\textbf{Intent} &
\textbf{Observed Operationalization} &
\textbf{Knowledge-Aware Operationalization} \\
\midrule
Comparative Grounding &
Unusually large withdrawals &
\texttt{WHERE amount > 10000} &
\texttt{WHERE amount > AVG(amount) + 2*STDDEV(amount)} \\

Role Confusion &
Most demanding jobs &
\texttt{ORDER BY COUNT(*)} &
Rank by qualification requirements \\

Process Reasoning &
Recurring overdrafts &
\texttt{WHERE balance < 0} &
Temporal persistence analysis over account history \\

Quantitative Reasoning &
Highest fatality rate &
\texttt{ORDER BY SUM(fatalities)} &
\texttt{SUM(fatalities) / COUNT(*)} \\

Policy Grounding &
Reportable transactions &
\texttt{WHERE amount > 50000} &
Apply policy-defined reporting threshold \\
\bottomrule
\end{tabular}
\end{table*}

\section{Recurring Failure Classes}

Across the 153 identified operationalization failures, we observed
five recurring classes summarized in Table~\ref{tab:failure_modes}. These categories are not intended as a complete
taxonomy, but rather recurring operationalization patterns
observed across domains and benchmarks.

\textbf{Comparative Grounding Failures} Comparative concepts such as \emph{high}, \emph{large},
\emph{unusual}, \emph{severe}, and \emph{extreme} require
reasoning relative to a reference population or distribution.
Although these concepts are expressed qualitatively in natural language, their operationalization typically depends on comparative reasoning over observed data. Generated workflows frequently operationalize such concepts using arbitrary fixed thresholds rather than contextual or distributional interpretations.
For example, \emph{unusually large withdrawals}
may be implemented using a fixed threshold without justification
that the threshold corresponds to unusual behavior within the
observed population.

\textbf{Role Confusion Failures} Many analytical concepts depend not only on an attribute's values but also on the analytical role the attribute plays within a computation. An attribute's role governs which operations are semantically meaningful or analytically admissible. Generated workflows frequently assign inappropriate roles to attributes, treating identifiers, frequencies, counts, or other observable quantities as direct evidence for higher-level concepts.
For
example, \emph{most demanding job titles} may be operationalized
as \texttt{ORDER BY COUNT(*)}, conflating job posting frequency
with qualification intensity. Similar failures arise when counts,
identifiers, or activity frequencies are used as proxies for concepts
such as importance, influence, risk, or complexity.

\textbf{Process-Level Failures} Many analytical intents describe processes unfolding over time
rather than properties of individual records. Meaningful states
often emerge from patterns across events rather than single
observations. Examples include recurring overdraft cycles,
worsening injury trends, sustained delinquency, and accelerating
transaction activity. Generated workflows frequently collapse such
process-oriented concepts into row-level predicates, ignoring
persistence, recurrence, trends, or temporal evolution. For
example, \emph{accounts with recurring overdraft cycles} may be
operationalized as a single negative balance rather than recurrence
across time.

\textbf{Quantitative Reasoning Failures} Many analytical requests require constructing derived quantities
through aggregation, normalization, or metric composition.
Generated workflows frequently ignore aggregation scope,
functional dependencies, or measurement-level constraints,
leading to semantically invalid derived metrics. For example,
\emph{total repayment} may be computed as
\texttt{amount + payments * duration}, double-counting principal
despite arithmetic validity. Similar failures arise when raw counts
are substituted for rates, incompatible quantities are combined,
or derived metrics are constructed without respecting their
underlying semantics.

\textbf{Policy Grounding Failures} Certain analytical concepts are defined by external policies,
regulations, or institutional standards rather than solely by
observed data. Generated workflows frequently operationalize
such concepts using arbitrary thresholds or local data statistics
while ignoring externally defined criteria. For example,
\emph{reportable incidents} may be implemented using a severity
threshold inferred from data distributions rather than the
governing policy definition.

Table~\ref{tab:operationalization_examples} illustrates how
operationalization failures can arise when the semantic information
required to express an analytical concept is unavailable to the
workflow-generation system.

\section{Implications for Agentic Data Systems}

Our findings suggest that correctly identifying a user's intent, grounding it to the appropriate schema elements, and generating an executable workflow are often not sufficient. Even when these steps succeed, the resulting computation may fail to faithfully capture the analytical concept expressed by the user.

One possible direction is to expose agents directly to additional sources of semantic information, including policies, documentation, process descriptions, and domain knowledge. However, reasoning over large collections of heterogeneous semantic artifacts introduces challenges in retrieval, grounding, consistency, and interpretation. 

An alternative direction is to develop intermediary semantic representations that explicitly capture operationalization knowledge in reusable form. Such representations could encode comparative baselines, process semantics, metric definitions, policy criteria, or admissible analytical operations, allowing agents to bridge the gap between analytical concepts and executable computations more systematically. Understanding what constitutes an admissible operationalization therefore represents an important direction for future research on agent-generated analytical workflows and AI-native data systems.

\section{Conclusion}

We presented a formative study of operationalization failures in
agent-generated analytical workflows. 
The recurring failure classes identified in our study suggest that
future agentic data systems may require richer semantic
representations capable of bridging the gap between user-level
analytical concepts and executable computations. Although our
experiments use a specific NL2SQL system as a concrete
instantiation, the failure classes identified in this
study concern the relationship between analytical concepts and
their computational operationalization rather than the behavior
of any single model.

An important direction for future work is therefore
to assess the extent to which the identified operationalization
failure classes persist across different workflow-generation
architectures, language models, and metadata environments.
Such studies would help distinguish failures arising from
limitations of particular systems from those reflecting more
fundamental challenges in bridging the semantic gap between
analytical concepts and executable computations. As AI-native systems increasingly generate analytical workflows
autonomously, understanding how analytical concepts should be
operationalized over available data represents an important
direction for future research.

\bibliographystyle{ACM-Reference-Format}
\bibliography{references}

\end{document}